\documentclass[a4paper,twocolumn,english,aps,prl,superscriptaddress]{revtex4-1}
\usepackage[T1]{fontenc}
\usepackage[latin9]{inputenc}
\usepackage{color}
\usepackage{babel}
\usepackage{amssymb}
\usepackage{graphicx}
\DeclareGraphicsExtensions{.pdf}
\usepackage{amsmath,bm,color}
\usepackage{float}
\usepackage{epstopdf}
\usepackage{tabularx}
\usepackage{braket}
\usepackage{booktabs}
\usepackage{array}
\usepackage{blindtext}
\usepackage{natbib}
\usepackage[unicode=true,pdfusetitle,
 bookmarks=true,bookmarksnumbered=false,bookmarksopen=false,
 breaklinks=false,pdfborder={0 0 1},backref=false,colorlinks=false]
 {hyperref}

\makeatletter

\usepackage{babel}
\usepackage{babel}

\begin{document}

\title{Altermagnetism and its induced higher-order topology on the Lieb lattice}

\author{Xingmin Huo}
\affiliation{School of Physics, Beihang University,
Beijing 100191, China}

\author{Xingchuan Zhu}
\affiliation{School of Intelligence Science and Technology, Nanjing University of Science and Technology, Nanjing 210094, China}

\author{Chang-An Li}
\affiliation{Institute for Theoretical Physics and Astrophysics, University of W$\ddot{u}$rzburg, 97074 W$\ddot{u}$rzburg, Germany}

\author{Shiping Feng}
\affiliation{Faculty of Arts and Sciences, Beijing Normal University, Zhuhai 519087, China, and School of Physics and Astronomy,  Beijing Normal University, Beijing 100875, China}

\author{Song-Bo~Zhang}
\affiliation{Hefei National Laboratory, Hefei, Anhui 230088, China}
\affiliation{International Center for Quantum Design of Functional Materials (ICQD), University of Science and Technology of China, Hefei, Anhui 230026, China}

\author{Shengyuan A. Yang}
\affiliation{Research Laboratory for Quantum Materials, Department of Applied Physics, The Hong Kong Polytechnic University, Kowloon, Hong Kong, China }

\author{Huaiming Guo}
\email{hmguo@buaa.edu.cn}
\affiliation{School of Physics, Beihang University,
Beijing 100191, China}

\begin{abstract}
Altermagnetism (AM) has brought renewed attention to the Lieb lattice. Here, we broaden the scope of altermagnetic models on the Lieb lattice by using a general scheme based on spin clusters. We design various altermagnetic models with $d$- and $g$-wave on the Lieb lattice, and investigate its interplay with spin-orbit coupling. While the altermagnetic unit cell reconstructs the topological edge states in the strip geometry and leads to the emergence of Dirac points, the in-plane magnetic moments of AM can induce gaps at these points. In an open square geometry, corner modes emerge within these gaps, realizing higher-order topological states. We further verify that the induction of higher-order topology is applicable to all altermagnetic configurations constructed here on the Lieb lattice, and is most pronounced for AM by comparing with the other types of magnetism such as ferromagnetism and ferrimagnetism. Our results highlight the exotic properties of AM, and suggest its potential applications in engineering topological quantum states.
\end{abstract}

\date{\today}

\maketitle
\section{Introduction}

Altermagnetism (AM) is a newly discovered magnetic phase characterized by alternating spin-split band structures and zero net magnetization~\cite{PhysRevX.12.031042,PhysRevX.12.040501,AdvFunctMater.2409327,jungwirth2024altermagnetsbeyondnodalmagneticallyordered,Tamang2025,npjSpintronics.3.1,jungwirth2024altermagnetsbeyondnodalmagneticallyordered,jungwirth2025altermagnetismunconventionalspinorderedphase}. 
The Lieb lattice underlies the crystal structure of a series of $d$-wave altermagnetic materials, including La$_2$CuO$_4$, La$_2$O$_3$Mn$_2$Se$_2$, KV$_2$Se$_2$O, and Rb$_{1-\delta}$V$_2$Te$_2$O, some of which have been experimentally confirmed using spin-resolved ARPES~\cite{Zeng2024,zhang2024crystalsymmetrypairedspinvalleylockinglayered,Nature2024Altermagnetism,Jiang2025,PhysRevLett.133.206401,NatCommun.15.2116,Nature.626.517,Nature.626.523,Nature.638.645,graham2025localprobeevidencesupporting,jaeschkeubiergo2025atomicaltermagnetism,PhysRevMaterials.9.024402,lin2024observationgiantspinsplitting,lin2024observationgiantspinsplitting,SciAdv.10.eadj4883,PhysRevMaterials.9.024402,Lee2024_BrokenKramersDegeneracy}. 
Accordingly, a two-dimensional minimal model on the Lieb lattice has been proposed to microscopically investigate altermagnetic properties~\cite{PhysRevB.108.224421,PhysRevLett.134.096703,PhysRevB.110.205140}. In this model, the magnetic moments adopt an inverse Lieb configuration, where nonmagnetic sites are surrounded by two magnetic sites with opposite spins. The two magnetic sites are related by a $C_4$ rotation symmetry, but not by inversion or lattice translation. As a result, both the spin-fermion and Heisenberg models exhibit spin-split electronic bands and chirally split magnon bands, which are hallmark features of a $d$-wave AM phase.

Numerous studies have focused on the AM on the Lieb lattice. The Lieb-lattice Hubbard model demonstrates the interaction-induced formation of altermagnetic Mott insulating ground states~\cite{syljuasen2025quantumgeometrymagnonhall,durrnagel2024altermagneticphasetransitionlieb,li2024dwavemagnetismcupratesoxygen,PhysRevB.108.L100402,PhysRevLett.132.263402,PhysRevLett.132.236701}. Superconducting instabilities from pairing interactions can be mediated by magnon, phonon, and antiferromagnetic fluctuations, leading to the realization of spin-polarized and spin-triplet superconductivity~\cite{,PhysRevB.109.134515,leraand2025phononmediatedspinpolarizedsuperconductivityaltermagnets,wu2025intraunitcellsingletpairingmediated,parthenios2025spinpairdensitywaves}. Furthermore, the Lieb-lattice AM can be tuned by strain and switched in a non-equilibrium manner through ultrafast optical methods~\cite{khodas2025tuningaltermagnetismstrain,eskandariasl2025controllingphotoexcitedelectronspinlightpolarization}.

The Lieb lattice hosts itinerant electrons with flat bands and high-spin Dirac fermions, making it a paradigmatic platform for exploring strong correlation effects and topological phenomena. The flat band, arising from the unequal number of sites in the two sublattices of the Lieb lattice, can support various correlated states due to enhanced interactions. When a Hubbard interaction is introduced, Lieb rigorously demonstrated that, at half-filling, the ground state exhibits a nonzero spin, serving as the first example of itinerant-electron ferromagnetism~\cite{PhysRevLett.62.1201}. Thereafter, the prototypes of correlated electrons, including the Hubbard model, the periodic Anderson model, and the Holstein model, have been continuously studied on the Lieb lattice using different many-body methods~\cite{,PhysRevB.106.224514,PhysRevB.96.245127,PhysRevB.101.165109,PhysRevB.106.205149,lima2025finitetemperaturesflatbands,PhysRevB.103.L220502,PhysRevB.98.094513,PhysRevA.96.053616,wei2025liebstheorembosehubbard,PhysRevLett.68.2648,PhysRevB.93.235143,PhysRevB.102.235152}. Recently, flat-band superconductivity in the attractive Lieb-lattice Hubbard model has begun to reveal the important role of the quantum metric in quantum matter~\cite{PhysRevB.90.094506,PhysRevB.106.014518,PhysRevB.111.L020506,CommsPhys.8.50,PhysRevResearch.2.023136}.

Pseudospin-1 Dirac fermions describe the low-energy excitations of itinerant electrons on a Lieb lattice, which generalize the spin-1/2 Dirac points in graphene~\cite{PhysRevB.81.041410}. Through spin-orbit coupling, a topological mass can be acquired, making Lieb lattice a host for a two-dimensional topological insulator~\cite{PhysRevB.82.085310,PhysRevB.97.081103,PhysRevB.100.235145,PhysRevB.96.205304,PhysRevB.99.125131,jiang2020topological}. The topologically nontrivial phase can even emerge dynamically through fermionic interactions~\cite{Tsai_2015,PhysRevB.100.045420}. The topological Lieb lattice, as a prototype for topological insulators, is actively engineered in artificial platforms such as circuit networks and optical lattices~\cite{NatPhys.13.668,PhysRevA.93.043611,NatureCommun.11.66,PhysRevB.109.054412,PhysRevB.97.075310,PhysRevA.83.063601,PhysRevB.99.235110}.

Motivated by the the rich physics hosted by the Lieb lattice, we investigate their intriguing interplays by combing different physical ingredients. 
In this study, we focus on the AM and the topological insulator on a Lieb lattice, and explore the quantum state induced by their coexistence, in particular the higher-order topological phases~\cite{benalcazar2017quantized,schindler2018higher,PhysRevB.99.245151}. We begin by performing an exhaustive design of altermagnetic models on the Lieb lattice, constructing various magnetic configurations that realize AM with $d$- and $g$-wave symmetries. We then incorporate these altermagnets into the topological Lieb lattice and find the emergence of higher-order topological states when the altermagnetic moments lie in the plane. Finally, by comparing with conventional magnetism, we confirm the significant effect of AM on generating higher-order topology on the Lieb lattice.

This paper is organized as follows. Section II introduces various altermagnetic models on the Lieb lattice, designed from spin clusters. Section III studies the interplay between spin-orbit coupling and AM. Section IV demonstrates the effect of AM on the Lieb lattice in inducing a higher-order topological phase. Section V examines the property of AM with other symmetries. Section VI presents the conclusions.

\section{Altermagnetic models on Lieb lattice}

\begin{figure}[hbpt]
  \includegraphics[width=7.5cm]{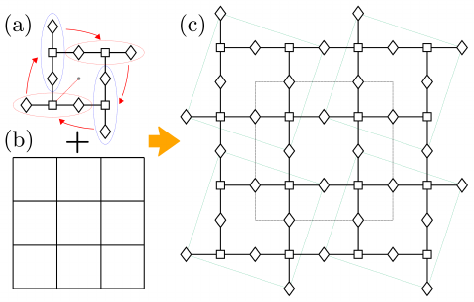}
  \caption{Schematic representation of the process of constructing a Lieb lattice using a three-site cluster. (a) The basis obtained by applying $C_{4z}$ to the three-site cluster. (b) The square Bravais lattice. (c) The Lieb lattice formed by attaching the basis from (a) to the square lattice in (b).}\label{fig1}
\end{figure}

\begin{figure}[hbpt]
  \includegraphics[width=8.5cm]{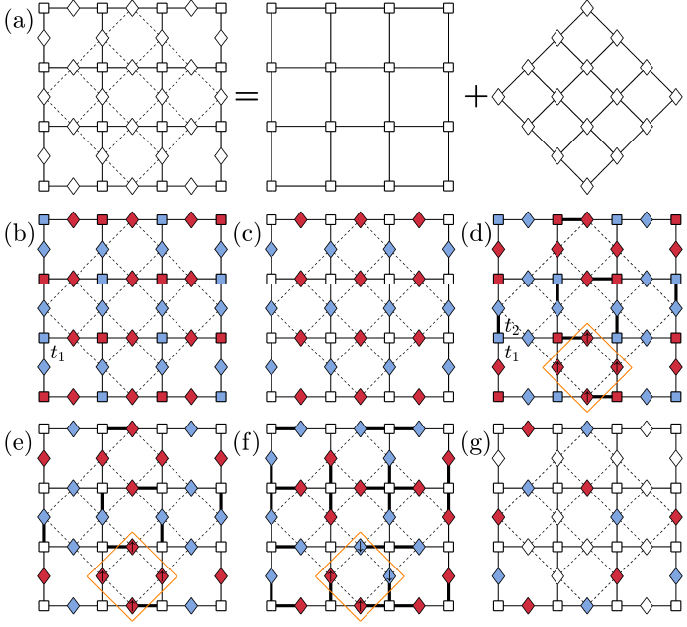}
  \caption{(a) Two sublattices of the bipartite Lieb lattice. (b)-(f) The constructed altermagnetic models, in which at least one sublattice is fully occupied. (g) A magnetic configurtion exhibitting $g$-wave AM. The thick bonds in (d) and (e) have a different hopping amplitude, $t_2$, resulting from the distortion introduced to break the symmetry that protects the spin degeneracy. All other bonds have a hopping amplitude of $t_1$. The $2\times2$ plaquette enclosed by the orange box is labeled as $(\uparrow,\uparrow,\uparrow,\uparrow)$ [(d) and (e)], and as $(\uparrow,\uparrow,\downarrow,\downarrow)$ [(f)].}\label{fig2}
\end{figure}

Previously, we proposed a general framework for generating altermagnetic models across various geometries from spin clusters~\cite{PhysRevLett.134.166701}. Here, we apply this approach to the Lieb lattice and conduct an exhaustive exploration of altermagnetic models on it. The Lieb lattice is characterized by the symmorphic crystallographic plane group $p4mm$, which allows for altermagnetic symmetries of $d$-wave or $g$-wave type. We first focus on the more common $d$-wave case, and subsequently construct $g$-wave altermagnetic models.

For $d$-wave AM, the key symmetry connecting the sublattices of opposite spins is $S = C_{4z}{\cal T}$, where ${\cal T}$ is time reversal operator and $C_{4z}$ is four fold rototion about $z$ axis. To ensure compatibility with the Lieb lattice, we select a spin cluster composed of three sites (each of which may be magnetic or non-magnetic) arranged in a row. We then apply $S$ to the spin cluster, generating a 12-site basis that respects $S$ [see Fig.~\ref{fig1}(a)]. Due to the symmetry constraint imposed by $S$, the local spins must be fully compensated within the basis. Since each site in the spin cluster can either be empty or occupied by an up-spin or a down-spin, excluding the empty cluster and clusters connected by spin flipping results in a total of 13 distinct spin clusters, corresponding to 13 different bases. To generate a Lieb lattice, we attach these bases to the square Bravais lattice, and construct the model Hamiltonian by adding on-site exchange and hopping terms. We consider the case of nearest-neighbor (NN) hopping, which is spin- and orientation-independent. 
It is noted that there is a many-to-one correspondence between the spin clusters and the resulting magnetic configurations. Spin clusters connected by an inversion operation with respect to the center will lead to the same configuration. Thus, in total, we obtain seven magnetic configurations that respect the symmetry $S$. Finally, we check whether the constructed Hamiltonians exhibit spin-split band structures. We find that three configurations still preserve the ${\cal PT}$ symmetry in real space protecting spin degeneracy of the band structure. Nevertheless, we can distort the basis to achieve spin splitting by introducing inhomogeneous hopping amplitudes while preserving the necessary four-fold rotational symmetry. After completing the above procedures, we construct six distinct types of altermagnetic models~\cite{note1}.

It is worth noting that we can also begin with a different choice of spin cluster containing four sites, resulting in eight types of altermagnetic models on the Lieb lattice (see Fig.\ref{afig2} in Appendix A for details). It is expected that larger spin clusters can be used, allowing for the construction of more altermagnetic models by the same procedure.


Figure \ref{fig2} illustrates several of the constructed altermagnetic models, in which at least one sublattice is occupied. These models contain fewer non-magnetic sites, making them potentially more relevant to realistic materials. Notably, the magnetic configurations in Fig.\ref{fig1}~(b) and (c) have been discussed in the context of $\text{CuO}_2$ plane of high-$T_c$ cuprates and identified as realizing AM. Specifically, the model in Fig.\ref{fig1}~(c) is widely studied in the literature as a minimal Hamiltonian to demonstrate the altermagnetic properties on the Lieb lattice~\cite{PhysRevB.108.224421}. The Lieb lattice is a bipartite lattice with a three-site unit cell, and can be decoupled into two sublattices, each arranged in a square geometry [see Fig.\ref{fig2}(a)]. The altermagnetic configuration can be understood in terms of the configuration in each individual sublattice. While the configuration in Fig.\ref{fig1}~(b) is a combination of two antiferromagnetic sublattices, the configuration in Fig.\ref{fig1}~(c) corresponds to antiferromagnetism occurring only on the tilted sublattice with more sites. Three of the newly identified altermagnetic models [Fig.\ref{fig2}(d)-(f)] extend the conventional antiferromagnetism on the sublattice with more sites to a $2 \times 2$ plaquette antiferromagnetism with different internal configurations: $(\uparrow,\uparrow,\uparrow,\uparrow)$ in Fig.\ref{fig2}(d) and (e), and $(\uparrow,\uparrow,\downarrow,\downarrow)$ in Fig.\ref{fig2}(f). All three magnetic configurations lead to $d_{x^2-y^2}$ AM.

\section{Interplay between spin-orbit couping and AM}

\begin{figure}[hbpt]
  \includegraphics[width=8.5cm]{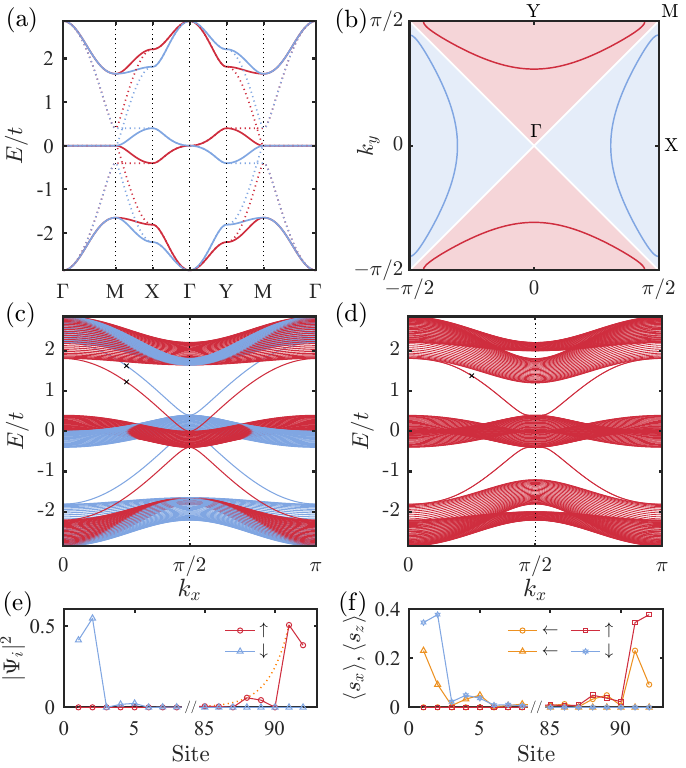}
  \caption{(a) The energy spectra of the Hamiltonian in Eq.(1) along the symmetric lines of the Brillouin zone for $\lambda=0$ (dotted line) and $\lambda=0.4$ (solid line). (b) The Fermi surfaces at $\mu=0.2$ for the AM with $\lambda=0$. The energy spectrum on a strip geometry with the magnetic moment in the $z$ direction (c) and in the $x$ direction (d). (e) Spatial distributions of the spin-up and spin-down edge states at $k_x=\pi/4$, indicated by the cross in (c), showing a monotonic decay with exponential modulation. {\color{blue}The red dotted line in (e) represents an exponential fit to the squared amplitudes on the same sublattice.} (f) Spin distributions $\langle s_x\rangle,\langle s_z\rangle$ of the edge states at $k_x=\pi/4$ in (d). The in-plane altermagnetism tilts the edge-state spins toward the $x$ direction, making them nonparallel. Here the nearest-neighbor hopping amplitude is $t=1$, and the exchange coupling in (c) and (d) is $J=0.4$.}\label{fig3}
\end{figure}

As a representative example, we then introduce the intrinsic spin-orbit (SO) interaction to the altermagnetic Hamiltonian in Fig.\ref{fig2}(c), which writes as 
\begin{align}\label{eq1}
H=&-t \sum_{\langle i j\rangle, \alpha} c_{i, \alpha}^{\dagger} c_{j, \alpha}+J \sum_{i, \alpha, \beta} \boldsymbol{S}_i \cdot c_{i, \alpha}^{\dagger} \boldsymbol{\sigma}_{\alpha \beta} c_{i, \beta}
\\ \nonumber
&+i \lambda \sum_{\langle\langle i j\rangle\rangle \alpha \beta}\sigma_{z, \alpha \beta} c_{i,\alpha}^{\dagger} c_{j,\beta}-\mu \sum_{i, \alpha} c_{i, \alpha}^{\dagger} c_{i, \alpha},
\end{align}
where $c_{i,\alpha}^{\dagger}$ and $c_{i,\alpha}$ are fermionic creation and annihilation operators with spin $\alpha,\beta\in\{\uparrow,\downarrow\}$ at site $i$. The notation $\langle ij \rangle$ represents nearest-neighbor pairs with a hopping amplitude $t$. The $J$ term describes the on-site exchange interaction between the local spin $\bm{S}_i$ and itinerant electron spin $\bm{\sigma}$. The chemical potential $\mu$ determines the electron filling of the system. 
$\lambda$ is the amplitude for the next-nearest-neighbor (NNN) spin-orbit-induced interaction, and $\boldsymbol{\sigma}=(\sigma_x,\sigma_y,\sigma_z)$ represent the Pauli spin matrices. Given that the spin-orbit term involves only $\sigma_z$ component, it preserves spin-$\sigma_z$ conservation. In the calculations, this \( J \)-term is treated within a mean-field approximation, 
where fluctuations of the local spins are neglected. The resulting frozen magnetic moments act as effective potentials for spin-up and spin-down electrons.

In momentum space, the Hamiltonian becomes $H$ $=\Sigma_{\mathbf{k} \sigma} \Psi_{\mathbf{k} \sigma}^{\dagger} \mathcal{H}_{\mathbf{k}} \Psi_{\mathbf{k} \sigma}$, where $\Psi_{\mathbf{k} \sigma}=\left(c_{A \mathbf{k} \sigma}, c_{B \mathbf{k} \sigma}, c_{C \mathbf{k} \sigma}\right)^T$ and $\mathcal{H}_{\mathbf{k}}=\mathcal{H}^0_{\mathbf{k}}\pm \mathcal{H}^{\lambda}_{\mathbf{k}}\pm \mathcal{H}_{J}$ [$+(-)$ for spin-up (spin-down)] with
\begin{align}\label{eq2}
\mathcal{H}^0_{\mathbf{k}}=\left(\begin{array}{ccc}
0 & -t(1+e^{-2ik_x}) & -t(1+e^{-2ik_y}) \\
& 0 & 0 \\
& & 0
\end{array}\right),
\end{align}

\begin{align}\label{eq2}
\mathcal{H}^{\lambda}_{\mathbf{k}}=\left(\begin{array}{ccc}
0 & 0 & 0 \\
& 0 & i\lambda(1-e^{2ik_x})(e^{-2ik_y}-1) \\
& & 0
\end{array}\right),
\end{align}
and $\mathcal{H}_{J}=\text{diag}(0,-m,m)$ with $m=JS$.
The lower triangle of the above matrices is understood to be filled so that the matrix is Hermitian.
The Hamiltonian matrix generally does not have a simple analytical form, we can obtain the polynomial equations satisfied by the eigenvalues,
\begin{align}
x^3-\left(16 \lambda^2c_{1, \bf k}+4 t^2c^+_{\bf k}+m^2\right)x\pm 4 m t^2c^-_{\bf k}=0,
\label{eq4}
\end{align}
where $c^+_{\bf k}=\cos ^2 k_x+\cos ^2 k_y$, $c^-_{\bf k}=\cos ^2 k_x-\cos ^2 k_y$, and $c_{1, \bf k}=\sin ^2 k_x \sin ^2 k_y$.

In the following two specific limits, we can derive the analytical expressions for the spectra.
When $m=0$ and $\lambda=0$, we obtain the spectrum of $\mathcal{H}_{\mathbf{k}}^0$ consisting of one degenerate flat band $E_{\mathbf{k}}^{(3)}=0$ and two dispersive bands,
\begin{align}
E_{\mathbf{k}}^{(1,2)}= \pm 2 t \sqrt{\cos ^2 k_x+\cos ^2 k_y}.
\end{align}
For $m=0$ and $\lambda\neq0$, the same degenerate flat band remains, while the dispersive bands are modified as,
\begin{align}
E_{\mathbf{k}}^{(1,2)}= \pm 2 \sqrt{t^2\left(\cos ^2 k_x+\cos ^2 k_y\right)+4 \lambda^2 \sin ^2 k_x \sin ^2 k_y},
\end{align}
when the system is a topological insulator with a gap  $\Delta_{\mathrm{SO}}=4|\lambda|$ opened at the Dirac point~\cite{PhysRevB.82.085310}. 

In the absence of SO interaction, the reduced Hamiltonian describes AM on the Lieb lattice.
It is clear that the polynomial equations for the eigenvalues in Eq. (\ref{eq4}) are spin dependent, thus the spin degeneracy will be lifted, which is verified by numerical solutions. Furthermore, the splitting between spin-up and spin-down depends on the crystal momentum $\mathbf{k}$, as can be seen from the spectrum along the high-symmetry lines in Fig.~\ref{fig3}(a). The energy spectrum exhibits spin-degenerate nodal lines along the diagonals of the Brillouin zone, and the spin polarization changes sign across the nodal line. This behavior corresponds to the spin-splitting characteristics of altermagnets with $d_{x^2-y^2}$-wave symmetry.
Figure \ref{fig3}(b) shows the Fermi surfaces of the electron dispersion. The spin-up and spin-down Fermi surfaces are anisotropic and mutually rotated by $90^{\circ}$, which further confirms that the magnetic system in Fig. \ref{fig2}(c) is a $d_{x^2-y^2}$-wave altermagnet.

The inclusion of SO interaction preserves the pattern of spin splitting, except that it opens gaps between adjacent bands. Along specific paths such as $\Gamma-X(Y)$, the energy spectrum is unaffected by the SO interaction, as the terms involving $\lambda$ in the polynomial equations in Eq. (\ref{eq4}) vanish.
We numerically solve the Hamiltonian $\mathcal{H}_{\mathbf{k}}$ in a strip geometry using direct diagonalization and find a pair of spin-filtered gapless states associated with each edge, as shown in Fig.~\ref{fig3}(c). Figure~\ref{fig3} (e) shows the probability distributions of the spin-filtered edge states, which demonstrate that, for each $k$, the two edge states carry opposite spins and are localized at opposite edges of the system. Moreover, they are exponentially localized from the boundaries into the bulk. These results indicate that the system, with both SO interaction and AM, remains in the topological phase, even though time-reversal symmetry is broken in the AM. Here, the magnetic moments in the altermagnetic system are aligned in the $z$-direction, thus the spin-up and spin-down subsystems are decoupled. The spectrum for the spin-up subsystem differs from that of the spin-down subsystem by a sign. The topological phase can be understood as persisting in the presence of an alternating on-site potential for each spin subsystem. Moreover, the system preserves a mirror symmetry $\mathcal{M}_z$, which allows the Hamiltonian to be block-diagonalized into mirror subspaces with eigenvalues $\pm i$. Each mirror sector carries an opposite Chern number ($1/3$ or $2/3$ filling), leading to a nonzero mirror Chern number $C_{\mathcal{M}}= \frac{1}{2}(C_{+i} - C_{-i})=1$ that characterizes a mirror Chern insulator protected by $\mathcal{M}_z$~\cite{PhysRevB.110.195142}. Since the eigenvalues of $\mathcal{M}_z$ correspond to the opposite spin sectors, the mirror symmetry effectively corresponds to $\sigma_z$ conservation, and the associated topological invariant is also referred to as the spin Chern number\cite{Sheng2006SpinChern}.

It is noted that the edge states exhibit spin splitting, similar to the case of ferromagnetism (FM). 
Furthermore, the magnitude of the splitting is precisely equal to $J$.
By examining the strip geometry, we observe that the magnetic moments at each outermost edge align in a ferromagnetic manner, which should account for the spin-dependent edge states. As a comparison, we also consider ferrimagnetism, where antiferromagnetic correlations exist between sites of the two sublattices, despite with the unequal number of sites. This leads to antiferromagnetic outermost edges, with the edge states being spin-degenerate.



We can also align the magnetic moments of the altermagnetism in the $xy$ plane. In this case, $\sigma_z$ is no longer a good quantum number, and the two spin sublattices become coupled. Supposing that the angle of the magnetic moments relative to the $x$-axis is $\phi$, the on-site exchange interaction is written as:
\begin{align}
H_J=JS \sum_{i, \alpha, \beta} c_{i, \alpha}^{\dagger} (\sigma_{x, \alpha \beta}\cos \phi+\sigma_{y, \alpha \beta}\sin \phi) c_{i, \beta}.
\label{eq3}
\end{align}
Since we find that the physical property is independent of $\phi$, we set $\phi=0$ in the following discussions, i.e., the magnetic moment is aligned along the $x$-axis.
Without AM, the topological edge states traverse the gap between the dispersive bands and the flat bands (with an $n$-fold degeneracy for a strip of width $3n + 2$). After the inclusion of magnetic moments, the degeneracy of the flat bands is lifted, similar to the case with perpendicular magnetic moments. However, in the presence of transverse magnetic moments, the edge states become degenerate, which can be understood using a perturbative method for small value of $J$. Under the basis $\Phi=(\phi_{\uparrow},\phi_{\downarrow})^{T}$ with $\phi_{\uparrow}=[c_{k_x\uparrow}^{(1,A)},c_{k_x\uparrow}^{(1,B)},c_{k_x\uparrow}^{(1,C)},...]^{T}$ and $\phi_{\downarrow}=[c_{k_x\downarrow}^{(1,A)},c_{k_x\downarrow}^{(1,B)},c_{k_x\downarrow}^{(1,C)},...]^{T}$ ( $i$ in $(i,\alpha)$ indexes the site and spans across the width of the strip), the Hamiltonian matrix for a strip geometry can be written as
\begin{align}
\mathcal{H}_{k_x}=\left(\begin{array}{cc}
\mathcal{H}^{t\lambda}_{\uparrow}(k_x) & \mathcal{H}_{J_x} \\
\mathcal{H}_{J_x}& \mathcal{H}^{t\lambda}_{\downarrow}(k_x)
\end{array}\right),
\end{align}
where $\mathcal{H}^{t\lambda}_{\alpha}(k_x)=\mathcal{H}^{0}_{\alpha}(k_x)+\mathcal{H}^{\lambda}_{\alpha}(k_x)$ originates from the hopping and the SO terms, while $\mathcal{H}_{J_x}=\text{diag}(0,JS,-JS,...)$ is a diagonal matrix coupling the opposite spin subsystems due to the $x$-direction exchange interaction.
At each $k_x$, there is an edge state $|\psi_{k_x}^{\alpha}\rangle$ corresponding to $\mathcal{H}^{t\lambda}_{\alpha}(k_x)$. While the spin-up and spin-down edge states are degenerate, their distributions are located near the opposite edges of the strip. Therefore, the average $\langle\psi_{k_x}^{\uparrow}|\mathcal{H}_{J_x}|\psi_{k_x}^{\downarrow}\rangle$ is nearly zero.
Treating the on-site exchange term $\mathcal{H}_{J_x}$ as a perturbation and under the subspace spanned by $|\psi_{k_x}^{\uparrow}\rangle$ and $|\psi_{k_x}^{\downarrow}\rangle$, the effective Hamiltonian can be found as
\begin{align}
\mathcal{H}_{eff}=\left(\begin{array}{cc}
\langle \psi_{k_x}^{\uparrow}|\mathcal{H}^{t\lambda}_{\uparrow}(k_x) |\psi_{k_x}^{\uparrow}\rangle& \langle \psi_{k_x}^{\uparrow}|\mathcal{H}_{J_x}|\psi_{k_x}^{\downarrow}\rangle \\
\langle \psi_{k_x}^{\downarrow}|\mathcal{H}_{J_x}|\psi_{k_x}^{\uparrow}\rangle& \langle \psi_{k_x}^{\downarrow}|\mathcal{H}^{t\lambda}_{\downarrow}(k_x)|\psi_{k_x}^{\downarrow}\rangle
\end{array}\right).
\end{align}
Since $\mathcal{H}_{eff}$ remains diagonal, the degeneracy of the edge states is unaffected by the AM order in $xy$ plane.

The persistence of edge states with the N\'eel vector aligned along the $x$-direction, as understood from above perturbative analysis, may stem from a bulk topological property protected by the combined symmetry $S = {\cal T}\hat{P}_{BC}$, where $\hat{P}_{BC}$ exchanges the spin-up and spin-down sublattices. However, the in-plane N\'eel vector tilts the spin-filtered edge states toward the $x$ direction [see Fig.~3(f)], making their spin orientations to become nonparallel and thereby rendering the edge states fragile even against nonmagnetic disorder that breaks $\hat{P}_{BC}$.

\section{Higher-order topology induced by AM on the Lieb lattice}

\begin{figure}[hbpt]
  \includegraphics[width=8.5cm]{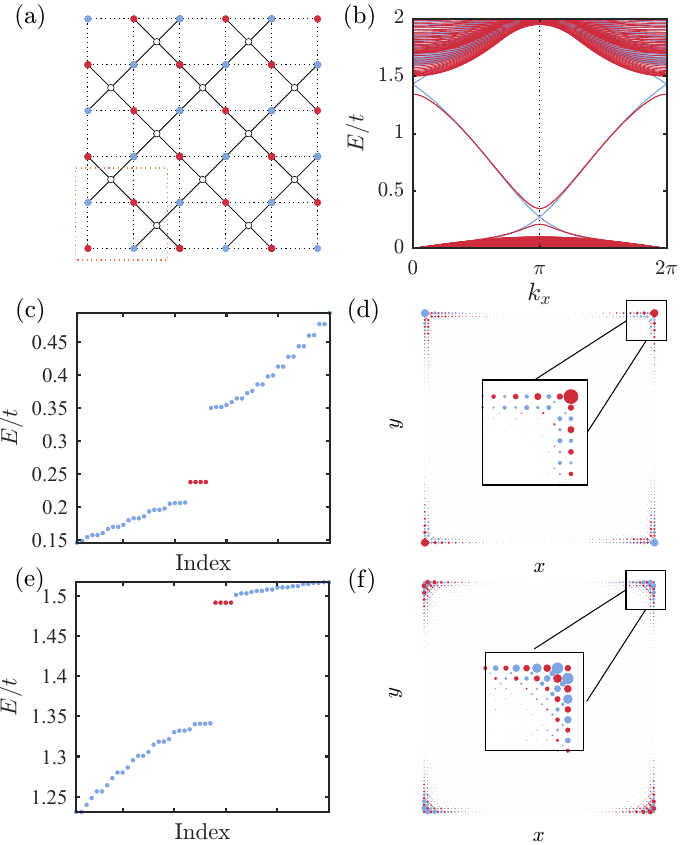}
  \caption{(a) Schematic of the Lieb lattice geometry with edges oriented in the $45^{\circ}$ direction. (b) The spectra of the corresponding strip geometry with (red) and without (blue) AM. (c) The spectrum of the open square geometry around the lower gap, showing four corner modes emerging in the gap. (d) The spin-resolved distributions of the corner modes near the Fermi energy. Due to the AM aligned in the $x$-direction, the spin is polarized in the $x$ (red) or $-x$ (blue) direction. (e) and (f) The spectrum around the upper gap and the spin-resolved distributions of the corner modes within the gap. Insets of (d) and (f) show a enlarged view of the corner. The parameters are $t = 1$, $\lambda = 0.4$, and $J=0.1$.}
  \label{fig4}
\end{figure}

Next, we consider an alternative strip geometry with the edges oriented in the $45^{\circ}$ direction, as shown in Fig.~\ref{fig4} (a). Here, we still focus on the altermagnetic configuration shown in Fig.~\ref{fig2}(c). 
Without AM, the topological edge states are reconstructed, leading to linear crossings at $k_x=0$ and $k_x=\pi$, respectively. AM in the $x$-direction opens gaps at the two momenta, causing the edge states to detach from the bulk bands and become floating within the band gap [Fig.\ref{fig4} (b)]. The floating edge states are known to be closely related with higher-order topological phases~\cite{PhysRevB.108.L100101,PhysRevB.107.174101,PhysRevResearch.4.023193,PhysRevB.111.045106}.
We then cut the strip into a square geometry with open boundary conditions, which preserves the combined symmetry of rotation and time reversal, $C_{4z} \mathcal{T}$. As shown in Fig.~\ref{fig4}(c), four states appear in each gap. Upon examining their distributions, these states are identified as corner modes. Therefore, AM in the $x$-direction induces higher-order topology in the tilted open square geometry. 

A general principle that a corner state will appear is the formation of a domain wall at the intersection of the two edges at the corner. In order to develop some intuition for these insulating phases, it is useful to examine the form of the low-energy Hamiltonians governing the excitations in the vicinity of the Dirac points. This is obtained by linearizing $\mathcal{H}_{k_x}=\mathcal{H}^0_{k_x}+\mathcal{H}^{\lambda}_{k_x}+\mathcal{H}_{J_x}$ near $k_x=0,\pi$ and subsequently projecting onto the subspace associated with the two bands in each edge. At $k_x=\pi$, the obtained effective Hamiltonian is,
\begin{align}
{\cal H}_{eff}^{AM}=\pm v k \sigma_z+JS(a\sigma_x+b\sigma_y)+c_0,
\label{heffam}
\end{align}
where $k$ denotes the momentum displacement from $k_x = \pi$; $v,a,b$ and $c_0$ are coefficients dependent on $t,\lambda$, with $v=0.285,a=0.6624,b=0.2275,c_0=0.2748$ for $t=1,\lambda=0.4$, and the sign $+(-)$ corresponds to the upper (lower) edge. The spectrum is given by $E^{\pm}=c_0\pm \sqrt{(v k)^2+(JS)^2(a^2+b^2)}$. Clearly, a gap of size $\sqrt{(JS)^2(a^2+b^2)}$ opens at $k=0$, (i.e., at $k_x=\pi$). The adjacent edges of the square geometry in Fig.~\ref{fig4}(a) have opposite exchange coupling $J$, leading to opposite mass terms in the Dirac Hamiltonian in Eq. (\ref{heffam}).
Hence, the mass necessarily undergoes a sign change across the corner. This soliton mass profile is known to produce massless states in the associated Dirac equation, which are localized near the corner, giving rise to the observed higher-order topological state.

In the vicinity of $k_x = 0$, the effective Hamiltonian retains the same form as that in Eq. (\ref{heffam}), although the coefficients differ: $v = -0.2448$, $a = 0.8211$, $b = -0.2707$, and $c_0 = 1.43$. It is noted that the magnitudes of the masses at $k_x = 0$ are larger than those at $k_x = \pi$, leading to relatively more localized corner states [see Fig.~\ref{fig4}(d) and (f)]. As $J$ increases, the corner states at $k_x=\pi$ become increasingly localized, while those at $k_x=0$ merge into bulk states.

The emergence of the higher-order topological phase is independent of the direction of the magnetic moment in the $xy$-plane. For a general magnetic moment $\vec{S} = (S_x, S_y)$ (the magnitude $S=\sqrt{S_x^2+S_y^2}$), the corresponding matrix in the projected space is $(aS_x - bS_y)J\sigma_x + (bS_x + aS_y)J\sigma_y$. Therefore, except for the renormalization of the coefficients of the two terms associated with exchange coupling, the form of the Dirac Hamiltonian and the spectrum remains unchanged, so does the higher-order topological phase. Moreover, in addition to the $45^{\circ}$ direction, other directions with nonzero angles also produce square geometries with AM-induced higher-order topology (see Appendix B).

\begin{table}
\begin{tabular}{|>{\centering\arraybackslash}p{1.5cm}|>{\centering\arraybackslash}p{2cm}|>{\centering\arraybackslash}p{2cm}|>{\centering\arraybackslash}p{2cm}|}
\hline
 &FM-1 &FM-2 & FIM \\
\hline
k$_x$=$\pi$ & a=0.00295 &a=0.0059& a=0\\
 &b=0.00859 &b=0.0172& b=0\\
\hline
k$_x$=0 &a=0.0322 &a=0.0643& a=0\\
 &b=-0.0976 &b=-0.195& b=0\\
\hline
\end{tabular}
\caption{The coefficients of the term associated with the exchange coupling $J$ in the projected effective Hamiltonian. FM-1(2) represent FM in the sublattice with more sites (the entire lattice). }\label{table1}
\end{table}

We have replaced AM with FM or ferrimagnetism FIM with the same exchange strength. It was found that while FIM does not open a gap in the edge spectrum, FM typically generates a much smaller gap than AM. This suggests AM as the most effective in inducing a higher-order topological phase among the magnetisms considered. This is evident in the coefficients of the term associated with the exchange coupling $J$ in the projected effective Hamiltonian, as shown in Table \ref{table1}.
Furthermore, by incorporating other forms of AM into the topological insulator on the Lieb lattice, we observe the emergence of apparent higher-order topological phase. Therefore, the property to induce higher-order topology in the topological insulator on the Lieb lattice appears to be most pronounced for AM.

It is noted that AM-induced higher-order topology has been studied using a heterojunction consisting of the Bernevig-Hughes-Zhang model and the effective Hamiltonian of AM \cite{PhysRevB.109.L201109, PhysRevB.109.245306, PhysRevLett.133.106601, PhysRevB.109.224502, PhysRevB.108.205410}. However, the specific details differ from the results based on topological Lieb lattice and altermagnetic configuration presented here. Depending on whether the AM is described by the $\mathbf{s} \cdot \hat{\mathbf{n}} \otimes \mathbb{I}$ or $\mathbf{s} \cdot \hat{\mathbf{n}} \otimes \tau_x$ channel (\( \mathbb{I} \) denotes the identity matrix and \( \tau_x \) is the \( x \)-component of the Pauli matrices), two or four corner modes may appear \cite{PhysRevB.109.L201109, PhysRevB.109.245306}. Furthermore, in the case of two corner modes, the positions of the corner states can be manipulated by adjusting the direction of the N\'eel vector in the $xy$ plane. Therefore, while AM can induce higher-order topological states, the specific results may vary depending on the forms used to describe both the topological insulator and AM.

\section{Higher-order topology induced by AM with other symmetries}

\begin{figure}[hbpt]
  \includegraphics[width=7.5cm]{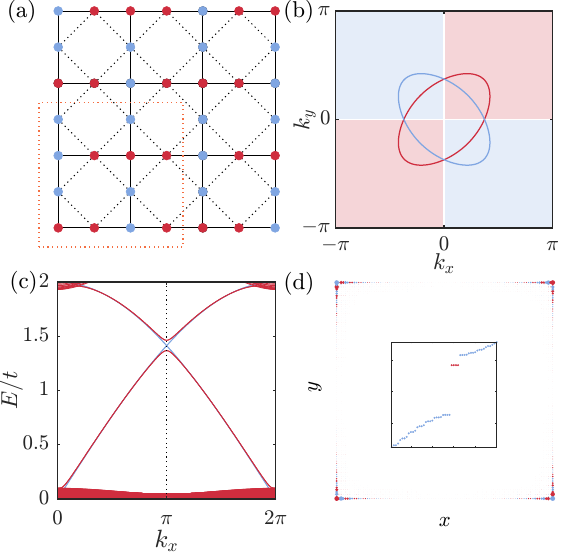}
  \caption{(a) Schematic of the $d_{xy}$-wave AM in a Lieb-lattice geometry with edges oriented in the normal ($0^{\circ}$) direction. (b) The Fermi surface for the altermagnetic model in (a) with $t = 1,J=0.4,\lambda = 0$, and $\mu=0.9$. (c) The spectrum of the corresponding strip geometry of (a), with (red) and without (blue) AM for comparison. (d) The distributions of the corner modes in the open square geometry of (a). Inset of (d) shows the corresponding spectrum, revealing four corner modes emerging in the gap. The exchange coupling is set to $J = 0.1$ in (c), and $J = 0.4$ in (d). The remaining parameters in (c) and (d) are $t = 1$ and $\lambda = 1$.}
  \label{fig5}
\end{figure}

Given the absence of higher-order topological states for specific orientations of the open square geometry, it is natural to ask whether the compatibility between the edge orientation and the spin-polarization pattern in momentum space is crucial for generating higher-order topology through AM. The question can be addressed by incorporating the $d_{xy}$ altermagnetic model in Fig. \ref{fig2}(a), where the pattern of the spin polarization is rotated by $45^\circ$ relative to that of the $d_{x^2-y^2}$ AM considered in the previous section [see Fig.\ref{fig5}(b)]. We identify AM-induced higher-order topological phases in square geometry with both $0^\circ$ and $45^\circ$ rotations. Figures \ref{fig5}(c) show the energy spectrum of a strip geometry with $0^\circ$ orientation. It is found that there appears one Dirac point in the edge dispersion without AM. After incorporating AM, the Dirac point becomes gapped, and corner modes emerge when the strip geometry is cut into an open square geometry.

In the case of a $45^\circ$ rotation, the results are similar to those of the $d_{x^2-y^2}$ AM: two Dirac points appear in the edge spectrum, and higher-order corner states exist within the two gaps opened by the $d_{xy}$ AM.
Therefore, for the same orientation between the edges and the spin polarization, higher-order topological states may or may not appear, implying that the orientation of the spin-polarization pattern is irrelevant for generating a higher-order topological phase. It is noted that the combined role of the geometry's orientation and the AM is to reshape or enlarge the unit cell, leading to the emergence of Dirac points at $J=0$ and their gap opening at $J \neq 0$, thereby realizing the higher-order topological insulator.

\begin{figure}[hbpt]
  \includegraphics[width=8.0cm]{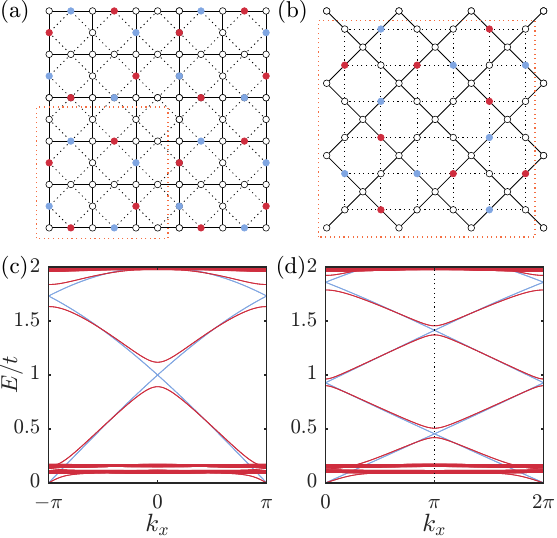}
  \caption{(a) Schematic of the $g$-wave AM in a Lieb-lattice geometry with edges oriented in (a) the normal ($0^{\circ}$) direction and (b) the $45^{\circ}$ direction. (c) and (d) Spectra of the corresponding strip geometries of (a) and (b), respectively, with (red) and without (blue) AM for comparison. The parameters in (c) and (d) are $t = 1$, $\lambda = 0.6$, and $J=0.4$.}
  \label{fig6}
\end{figure}

We also investigate the interplay between the $g$-wave AM [see Fig.~\ref{fig6}] and SO coupling on the Lieb lattice. It is worth noting that while previous studies have primarily focused on $d$-wave altermagnetic models on the Lieb lattice, the $g$-wave case is explored here for the first time. We find that the effect of $g$-wave AM is similar to that of the $d_{xy}$-wave AM described above, with higher-order topological states appearing in both geometries with normal and $45^{\circ}$ orientations. Since the $g$-wave altermagnetic model has a significantly larger unit cell, multiple edge-band foldings generate two Dirac points and four Dirac points in the $0^{\circ}$-oriented geometry and $45^{\circ}$-oriented geometry, respectively. Consequently, higher-order corner modes emerge at all these locations in the spectra of open squares.

\section{Conclusions and discussions}
We conduct a systematic study of AM and its induced higher-order topological states on the Lieb lattice. Using the general framework for constructing altermagnetic models proposed in Ref.\cite{PhysRevLett.134.166701}, we obtain various magnetic configurations on Lieb lattice that can support $d_{x^2-y^2}$-, $d_{xy}$-, and $g$-wave AM.
We then incorporate spin-orbit coupling into these altermagnetic models and find that the bands exhibit features of both alternating spin splitting and topological gaps, indicating the coexistence of altermagnetic and topological phases. When the magnetic moments lie in-plane, the AM can open a gap at the Dirac points in the spectrum of the strip geometry. Subsequently, in the corresponding open square geometry, four modes appear within each gap, with their distribution localized near the corners, thereby realizing the higher-order topological insulator. 
In contrast, conventional magnetism, such as ferromagnetism and ferrimagnetism, opens a much smaller higher-order topological gap, highlighting the pronounced effect of AM in inducing a higher-order topological phase on Lieb lattice. Our results enhance the understanding of magnetism-induced higher-order topology on Lieb lattice, which may help in engineering exotic topological quantum states with AM and in identifying AM through heterostructures composed of topological insulators and AM.

\begin{acknowledgments}
The authors thank Z.-B. Yan, L.-H. Hu, Z.-Q. Liu, Y. Du, C.-C. Liu, Q.-H. Liu, Y.-T. Yu for
helpful discussions. X.H. and H.G. acknowledges support from the NSFC grant No.~ 12574249 and the BNLCMP open research fund under Grant No. 2024BNLCMPKF023. X.Z. acknowledges support from the Natural Science Foundation of Jiangsu Province under Grant BK20230907 and the NSFC grant No.~12304177. C.L. thanks the support from W$\ddot{\text{u}}$rzburg-Dresden Cluster of Excellence ct.qmat, EXC2147, project-id 390858490. S.F. is supported by the National Key Research and Development Program of China under Grant Nos. 2023YFA1406500 and 2021YFA1401803, and NSFC under Grant No. 12274036. S.Y. acknowledges support from UM MYRG (GRG2024-00018-IAPME). S.B.Z. was supported by the start-up fund at HFNL, the Innovation Program for Quantum Science and Technology (Grant No. 2021ZD0302800) and the National Natural Science Foundation of China (Grant No. 12488101).
\end{acknowledgments}


\appendix
\setcounter{equation}{0}
\renewcommand\theequation{A.\arabic{equation}}
\setcounter{figure}{0}
\renewcommand{\thefigure}{A\arabic{figure}}

\section{More altermagnetic models on Lieb lattice}

\begin{figure}[hbpt]
  \includegraphics[width=7.5cm]{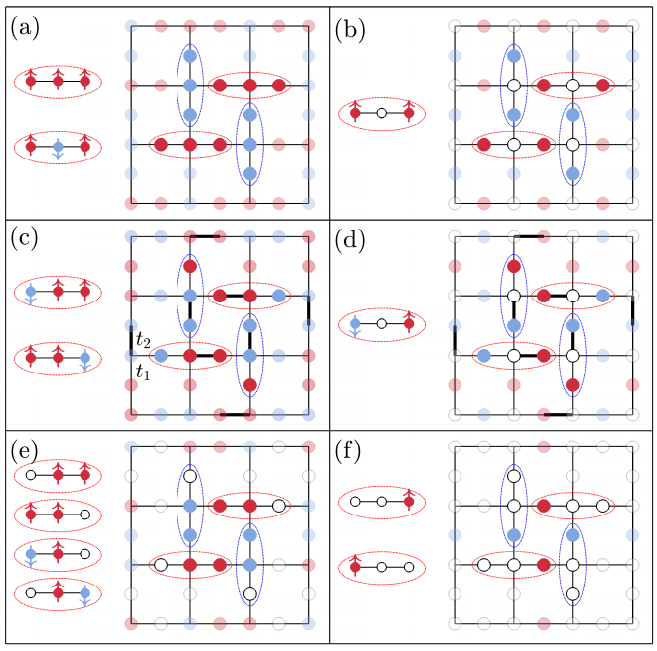}
  \caption{Six types of altermagnetic models on the Lieb lattice designed using a three-site spin cluster. The left part of each panel shows the starting spin cluster. All possible combinations in which each site within the spin cluster can be occupied by either an empty, up-spin, or down-spin state are considered, except for the one with only the central site occupied, in which the spin degeneracy cannot be lifted by a simple distortion. All bonds have a hopping amplitude $t_1$, except for the thick bonds in (c) and (d), which result from the necessary distortion and have a hopping amplitude $t_2$.}\label{afig1}
\end{figure}

\begin{figure}[hbpt]
  \includegraphics[width=7.5cm]{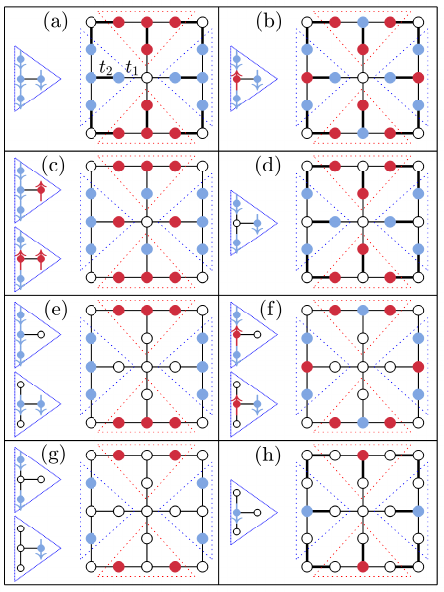}
  \caption{Eight types of altermagnetic models on the Lieb lattice designed using a four-site spin cluster. The left part of each panel shows the starting spin cluster. All bonds have a hopping amplitude $t_1$, except for the thick bonds in (a), (b), (d) and (h), which result from the necessary distortion and have a hopping amplitude $t_2$.}\label{afig2}
\end{figure}

Here, we demonstrate all altermagnetic models on the Lieb lattice generated using three-site and four-site spin clusters in Figs.~\ref{afig1} and~\ref{afig2}.
The characteristic spin splitting of these identified altermagnets has been verified by diagonalizing the following altermagnetic Hamiltonian, 
\begin{align}
H=&-t_1 \sum_{\langle i j\rangle_1, \sigma} c_{i, \sigma}^{\dagger} c_{j, \sigma}-t_2 \sum_{\langle i j\rangle_2, \sigma} c_{i, \sigma}^{\dagger} c_{j, \sigma} \nonumber
\\
&-J \sum_{i, \sigma, \sigma^{\prime}} \boldsymbol{S}_i \cdot c_{i, \sigma}^{\dagger} \boldsymbol{\sigma}_{\sigma \sigma^{\prime}} c_{i, \sigma^{\prime}}-\mu \sum_{i, \sigma} c_{i, \sigma}^{\dagger} c_{i, \sigma}.
\label{aeq1}
\end{align}

\begin{figure}[hbpt]
  \includegraphics[width=7.5cm]{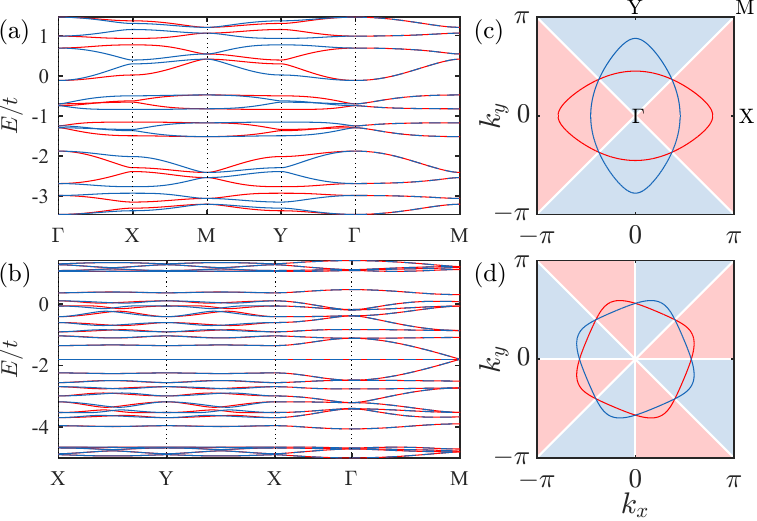}
  \caption{The energy spectra along the symmetric lines of Brillouin zone (left) and the Fermi surfaces (right). (a) and (b) correspond to the $d$-wave altermagnetic Hamiltonian in Fig.\ref{fig2}(e) with the parameters $t_{1}=1$, $t_{2} = 0.3$, $m=0.6$ and $\mu = 1$. (c) and (d) correspond to the $g$-wave altermagnetic Hamiltonian in Fig.\ref{fig2}(g) with the parameters $t_{1}=1$, $m=2$ and $\mu = 1.8$.}\label{afig3}
\end{figure}

Figure \ref{afig3} (a) shows the spectrum along the high-symmetry lines of the altermagnetic models in Fig.\ref{fig2}(e), 
which demonstrates the spin-splitting characteristics with a $d_{x^2-y^2}$-wave symmetry. The spin-up and spin-down Fermi surfaces are anisotropic and mutually rotated by $90^{\circ}$, which further confirms that the magnetic system in Fig.\ref{fig2}(e) is a $d_{x^2-y^2}$-wave altermagnet. 
In Fig.~\ref{afig3} (b) and (d), we also plot the spin-split spectrum and Fermi surface for the altermagnetic model in Fig.~\ref{fig2}(g), which demonstrates its $g$-wave spin-splitting characteristics in the Brillouin zone.  

It should be noted that the altermagnetic models in Figs. \ref{afig1} and \ref{afig2} all exhibit $d_{x^2-y^2}$-wave symmetry, except for the one in Figs. \ref{afig1}(a), which is a $d_{xy}$-wave AM.

\section{Other open square geometries with different oriented edges}

\begin{figure}[hbpt]
  \includegraphics[width=7.5cm]{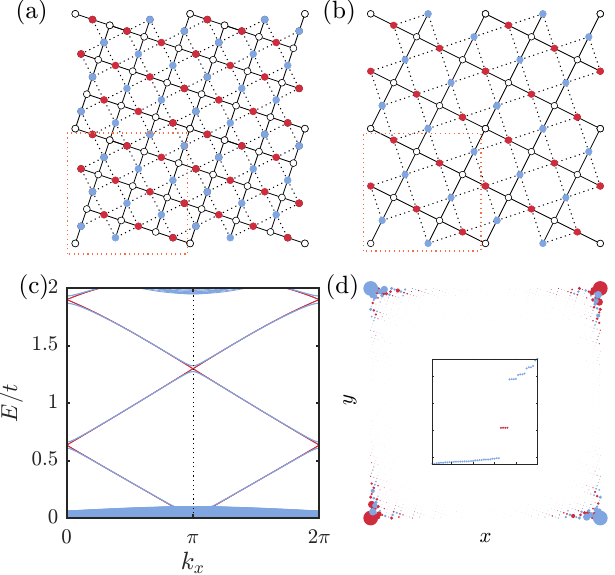}
  \caption{Open square geometries with edges oriented at angles $\theta_1=\text{atan}(1/3)\approx 18.43^\circ$ (a) and $\theta_2=\text{atan}(1/2)\approx 26.57^\circ$ (b) directions. The AM in Fig.\ref{fig2}(c) can induce higher-order topological states in both geometries. (c) The spectra of the corresponding strip geometries of (a) with (blue) and
without (red) AM for comparison. (d) The distribution of the corner mode of the open square geometry in (a), with the corresponding spectrum shown in the inset. The exchange couplings are $J=0.1$ in (c) and $J=0.4$ in (d). The other parameters are $t=1$ and $\lambda=0.4$.}\label{afig4}
\end{figure}

For the AM in Fig.~\ref{fig2}~(c), we investigate the induced higher-order topological states in open square geometries with differently oriented edges, specifically $\theta_1 = \text{atan}(1/3) \approx 18.43^\circ$ and $\theta_2 = \text{atan}(1/2) \approx 26.57^\circ$ [see Fig.~\ref{afig4}(a) and \ref{afig4}(b)]. It is found that the alternmagnetism in Fig.~\ref{fig2}(c) can induce higher-order topological states in both geometries. As shown in Fig.~\ref{afig4}(c), the spectra of the corresponding strip geometries in Fig.~\ref{afig4}(a) are similar to the case with a $45^\circ$ oriented edge. Accordingly, alternmagnetism can induce higher-order topological states in the open square geometry in Fig.~\ref{afig4}(a). The situation for the geometry in Fig.~\ref{afig4}(b) is similar, and thus the detailed results are not shown here.

\section{Spin and Magnetic Space Group of Altermagnetic Configurations on the Lieb Lattice}

To facilitate the identification of candidate materials relevant to our models, we list the spin and magnetic space groups of the identified altermagnetic configurations on the Lieb lattice\cite{PhysRevX.14.031038}.

\begin{table}
\begin{tabular}{|>{\centering\arraybackslash}p{2.5cm}|>{\centering\arraybackslash}p{3cm}|>{\centering\arraybackslash}p{2cm}|}
\hline
 Configuration &SSG &MSG \\
\hline
Fig. 2(b) & P$^{-1}$4/$^1$m$^{-1}$b$^{1}$m$^{\infty m}$1 &P4$'$/mbm$'$\\
\hline
Fig. 2(c)  &&\\
Fig. 2(f) & &  \\
Fig. A1(f) &P$^{-1}$4/$^1$m$^{1}$m$^{-1}$m$^{\infty m}1$  &P4$'$/mm$'$m  \\
Fig. A2(a)-(c) & & \\
Fig. A2(e)-(h)& & \\
\hline
Fig. 2(d)-(e) &P$^{-1}$4/$^1$m$^{\infty m}1$ &P4$'$/m\\
Fig. A1(e) & & \\
\hline
Fig. 2(g) &P$^{1}$4/$^1$m$^{-1}$m$^{-1}$m$^{\infty m}1$ &P4/mmm\\
\hline
\end{tabular}
\caption{Spin space group (SSG) and magnetic space group (MSG) of the altermagnetic configurations on the Lieb lattice. The data is obtained using online program "FINDSPINGROUP".  }\label{table1}
\end{table}

\bibliographystyle{apsrev4-1}
\bibliography{ref}

\end{document}